# Revisiting the scientometric conceptualization of impact and its measurement[1]

*Giovanni Abramo*
Laboratory for Studies in Research Evaluation
Institute for System Analysis and Computer Science (IASI-CNR)
National Research Council of Italy
ADDRESS:   Istituto di Analisi dei Sistemi ed Informatica (IASI-CNR)
Via dei Taurini 19, 00185 Roma - ITALY
tel. and fax +39 06 72597362, giovanni.abramo@uniroma2.it

**Abstract**
The development of scientometric indicators and methods for evaluative purposes, requires a multitude of assumptions, conventions, limitations, and caveats. Given this, we cannot permit ambiguities in the key concepts forming the basis of scientometric science itself, or research assessment exercises would rest on quicksand. This conceptual work attempts to spell out some principles leading to a clear definition of "impact" of research, and above all, of the appropriate scientometric indicator to measure it. The aim is to stimulate a discussion aimed at a definitive convergence on the meaning and measurement of a fundamental concept of the scientometric science.

**Acknowledgements**

My heartfelt thank goes to David Pendlebury, who is to me a living encyclopedia of scientometrics. His comments and insights made me aware of how poor was my first version of this manuscript. I am also sincerely grateful to Ludo Waltman for the fruitful discussion and suggestions which elicited significant improvements in the work. Last but not least, I wish to thank Lutz Bornmann and Cassidy Rose Sugimoto, for reviewing an earlier version of this manuscript. It must be noted that their support is not to be interpreted as a full alignment with the thoughts and perspectives presented in the paper, of which the author assumes sole responsibility.

---

[1] Abramo, G. (2018). Revisiting the scientometric conceptualization of impact and its measurement. *Journal of Informetrics*, 12(3), 590-597. DOI: 10.1016/j.joi.2018.05.001

# 1. Introduction

"Impact", jointly with "citation", is probably the most frequent term in scientometric literature. It recurs so often that its definition is more and more taken for granted, hardly ever spelled out. Impact is often, and much more easily, defined by means of the indicators employed to measure it, rather than as a concept in itself. It is precisely the proliferation of measurement indicators, witnessed by all, that calls for a red flag and return to home base. Without this, there is serious risk that we will lose sight of what we are measuring and, above all, of what we want to measure. The obsessive pursuit of measurement – "measure all that is countable, regardless of whether it counts or not" – resembles a case of gold fever. One apparent aspect of this, which calls for remediation, is to "first measure, and then ask whether it counts". An occurrence would be the situation of the altmetric indicators, which have proliferated in the past few years. First they were constructed, and now scientometricians strive to find out what they actually measure and how they might be used (Priem, & Costello, 2010; Thelwall, Tsou, Weingart, Holmberg, & Haustein, 2013; Fenner, 2014; Haustein, Larivière, Thelwall, Amyot, & Peters, 2014; Costas, Zahedi, & Wouters, 2015; Thelwall, 2017; Sugimoto, Work, Larivière, & Haustein, 2017).

The problem I see, and the point I am making, is that there cannot be different definitions of a concept that one intends to measure, nor a myriad of relevant measurement indicators. A number of scholars and organizations have advised that a clear definition of impact is required (Duryea, Hochman, and Parfitt, 2007; Grant et al., 2009; Russell Group, 2009). The aim of the current work then is to revisit the scientometric conceptualization of impact and the relevant measurement indicator. The hope is that we will begin to clear the fog over the meaning and the measurement of a concept that lies at the heart of evaluative scientometric studies. A clear definition of the concept and its measurement (and the limits embedded) would also help to blunt weapons of the main detractors of evaluative scientometrics. Last but not least, if a convergence on the meaning of research impact and method of measurement is achieved, then scientometric studies by different researchers may be compared more accurately.

It is important to note that the argument presented in this paper depends on the perspective that one has on the definition and scope of the field of scientometrics. A number of scholars broadly define scientometrics as the quantitative study of science, implying that it can cover some of the work done in the economics of science. My own perspective reflects the current reality, whereby scientometrics is contiguous to such fields as the economics of science and the economics of innovation. Although disciplines generally present blurring boundaries and some overlapping domains, my perspective is that scientometrics does not embed the above two fields, as it lacks the "socio-economics metrics".

The following questions drive the work:
1. What do (and should) scientometricians mean by "impact" of research?
2. How then should scientometricians measure it, given the tools they have available?

The search for answers should sharpen our image of what it is we want to measure, what we can actually measure, and above all, what we do not succeed in measuring.

# 2. The conceptualization of impact of research

## 2.1 A review of the literature

In reviewing the definitions of scientometrics in the literature, Mingers and Leydesdorff (2015) find that the main themes of scientometric research "include ways of measuring the quality and impact of research". The authors distinguish quality from impact of research. I will revisit this distinction at a later stage. Talking about impact, they identify that scientometrics deals with the impact "of some thing" (research), but not "on what", leaving the latter implicit. In the scientometric literature, in addition to the "research", we often find impact associated with terms that are somewhat different: publications, journals, individual researchers, research groups, institutions and nations. In some cases, the question of "on what" can be inferred from a modifier, identifying the scope of the root word: e.g. as in "scientific, academic, scholarly" impact, "non-scholarly, educational" impact, or "economic, social (societal)" impact. "Impact" is also often modified by terms denoting the means of measurement, indicating that we in fact use scientometric tools: "bibliometric, scientometric" impact or "citation" impact.

In his book "Citation analysis in research evaluation", Henk Moed (2005, p. 221) states: "The term *impact*, coined by Garfield and later used as a key concept by Martin and Irvine, is often used, but it is suggested to use the term *citation impact*, as it expresses the methodology along which impact is measured". It is interesting to recall then what Eugene Garfield, the father of scientometrics, meant by impact. David Pendlebury, one of the earliest and closest collaborators of Dr Garfield in personal research projects, interviewed by me on the issue, states that Garfield's conceptualization of impact is "a little loose, originally, but he mentions *impact on the literature* in his 1955 paper in Science (Garfield, 1955), and then in 1963 Irving Sher and he address impact in terms of a size-independent journal measure (Garfield & Sher, 1963)". Pendlebury further adds: "And I do admit that in the internal jargon of ISI[2] over the years, we understood "impact" to mean a size-independent measure of research performance (not productivity), in other words, citations divided by publications. This is the batting average concept. So our own definition evolved from Gene's early conception of citations as indicators of influence on the literature (and how the literature changed because of select papers, journals, etc.) to a narrow definition of a class of citation indicator (size independent). But we didn't go beyond that, by which I mean we advanced no theory or philosophy of *research impact*".

In their seminal work on assessing basic research, Martin and Irvine (1983) state that "the impact of a publication describes its *actual* influence on surrounding research activities at a given time", and further on "it is the impact of a publication that is most closely linked to the notion of scientific progress - a paper creating a great impact represents a major contribution to knowledge *at that time*". Along the same lines, Moravcsik (1977) holds that the impact of a publication lies in its influence on subsequent publications, and that such influence will manifest itself by the influenced paper citing the influencing paper.

More recently, we have witnessed a tendency to broaden the definition of research impact (Bornmann, 2014), including societal and economic impacts, and a growing pressure on scientometricians, by governments and funding agencies, to measure them. Through time, a number of scholars and practitioners have proposed their own perspectives on the meaning of impact, ending up extending the original definition and leading to the current situation where "impact" is often used in vague, generic ways; stakeholders and organizations use different definitions and think differently; they often talk past one another; review articles on impact definitions and relevant measurement have been published. Examples are the review by Penfield, Baker, Scoble, and Wykes (2014) which summarizes a wide range of definitions of impact, and Bornmann's work (2017) describing how impact is generally measured. I refer the reader to the above reviews for additional concrete examples. A convergence on the basic meaning of impact, and how to measure it is now strongly needed.

---

[2] The Institute for Scientific Information (ISI) was founded by Garfield in 1960.

Going back to the supposed distinction between "quality" and "impact", in addition to Mingers and Leydesdorff (2015) a number of other scholars distinguish between them (Cole & Cole, 1973; Martin & Irvine, 1983; [3] Sugimoto & Larivière, 2018, p. 66). A review work by Leydesdorff, Bornmann, Comins, and Milojević (2016) summarizes the perspectives of a number of scholars on the difference between quality and impact. Most often, definitions are simply stated without any supporting arguments. To exemplify, according to Martin and Irvine (1983), "quality is a property of the publication and the research described in it. It describes how well the research has been done, whether it is free from obvious "error", how aesthetically pleasing the mathematical formulations are, how original the conclusions are, and so on". Further, quality "is not just intrinsic to the research, but is something judged by others who, with differing research interests and social and political goals, may not place the same estimates on the quality of a given paper. Even the same individual may evaluate the quality of a paper differently at different times because of progress in scientific knowledge and shifts in his or her location." As the authors themselves state, the above definition of quality is highly subjective.

Contrary to what most believe quality does not need to be a subjective attribute. According to the ISO 9000: 2015 International Standard, which provides the fundamental concepts, principles and vocabulary for quality management systems (QMS) and the foundation for other QMS standards, the technical definition of quality is "the degree to which a set of inherent characteristics fulfils a requirement".[4] Given that in general the ultimate requirement of a publication is that it provides impact on future scientific advancement, quality needs to refer to impact, and the measure of quality and impact would then be synonymous.

**2.2 The role of scientometrics in assessing impact**

To better appreciate the role of scientometrics in assessing impact, we need to widen our horizons, observing its position within the overall research value chain, which starts with planning and funding, proceeds with project selection and execution, and ends with the research results, which once transferred (used), produce outcomes. Scientometrics is grounded in the quantitative analysis of scientific advances, mainly in the area of the "research results", for which it tries to measure impact, for evaluative purposes. Funders, be they governmental or private sector, must periodically decide how much of their budgets to devote to research, how to divide the funds among research programs and disciplines, and how (on competitive and/or non-competitive bases) to assign these allocations to research institutions, individuals or businesses. It is the policy makers and top managers who make such decisions, based on policy goals, strategic analyses and financial constraints. Incidentally, because scientometric analysis can be used to assess strengths and weaknesses at the discipline, institution and territorial level, such analysis, correctly applied, can contribute to informing the various strategic decisions. However, the essential characteristic of the value chain is that the funders share one long-term (grand) goal: "the maximization of returns on research investment". For the public sector this would be explicitly stated as social returns, but the same qualification increasingly holds true for the private sector. Although the private sector is traditionally considered as interested only in maximizing financial and strategic payoffs from research investments, the growing pressure for corporate social responsibility (CSR) has led managers to initiate assessments, and to take responsibility for their companies' effects on social wellbeing (Wood, 1991).

---

[3] More precisely, Martin and Irvine (1983) argue that "it is necessary to distinguish between, not two, but three concepts - the "quality", "importance", and "impact" of the research described in a paper", whereby "importance" of a publication refers to its *potential* influence on the advance of scientific knowledge.

[4] ISO (the International Organization for Standardization) is a worldwide federation of national standards bodies. ISO 9000: 2015 is accessible at https://www.iso.org/obp/ui/#iso:std:iso:9000:ed-4:v1:en, last accessed 8 April, 2018.

Whenever scientometricians assess impact they must then recall that in general, whatever the provisional or specific purposes, the objectives of the assessment are, at least indirectly, to contribute to the above grand goal. We can then agree with Mingers and Leydesdorff (2015) that assessments should measure the impact of research, and I would add that the ultimate goal is to assess "impact of research on society". The question becomes: "does scientometrics in fact measure the research impact on society, leading to calculations of social return on research investment?" The answer is no, as seen by examining both the current position of scientometrics in the value chain, where it concentrates on evaluating the "impact of research results", with rare references to any prior and subsequent stages, and the tools at its disposal: the biblio-metrics of science notably differ from the metrics applied in economics. The current scope of scientometrics in fact is too narrow to encompass the overall chain of value added activities, from research funding to social impact.

Governments and private organizations invest in research to accomplish scientific and technical advancement,[5] which has the potential of contributing to socio-economic progress, or returns. The Social Return on Research Investment (SRORI) can then be thought of as the product of two factors, expressed in formulae:

SRORI = Research Productivity · Research Yield

where the first factor, research productivity, is the ratio of scientific advancement to money invested; and the second factor, research yield, is the ratio of social return to the scientific advancement.

Given a scientific or technical advancement, its research yield is a function of the effectiveness and timeliness of: its incorporation in technologies (vertical technology transfer); their industrialization, or integration in operations, products and services; the resulting commercialization; market dynamics; and finally, the environment. The measurement of the final yield from research is beyond the scope of scientometrics, particularly given the current limitations in tools. In fact scientometrics deals only with the inputs and outputs of research (more precisely, with tangible knowledge encoded in documents of different types, and indexed in bibliometric repositories), and the embedded references to previous research, but does not deal with the assessment of social impact[6] from those outputs, nor with the monetization of the financial and extra-financial value of outcomes (externalities, economic surplus, etc.). The latter is dealt with in the domain of economics of science and innovation. Tracing the flow of scientific output to technology, leaving aside the question of the latter's social impact, is a formidable task per se, implying a long time horizon. Project Hindsight was the first of a series of attempts of this kind (Sherwin & Isenson, 1967): over a 20-year horizon, very little contribution from basic science was found. Project TRACES (IIT Research Institute, 1968) succeeded in tracking the origins of six technological innovations back to the underlying basic sciences, but only after extending the time horizons well beyond twenty years. Other projects followed in the pioneering years of scientometrics (Langrish, 1971; Gibbons & Johnston, 1974), but apart from limited variations, the evidence was that the normal path from science to technology requires a great amount of time – not considering the time from technology to commercialization, and the product-life cycles along which social benefits might accumulate. Given the need for timeliness in research evaluation, scientometrics instead deals in short time horizons.

---

[5] We leave aside here arts and humanities research, whose impact requires an *ad hoc* definition.

### 2.3 A definition of research impact and relevant measurement indicator

Having clarified that the research impact measured by scientometrics is not social impact, then what kind of impact is it? To answer this question we have to return to the very essence of scientific activity, which is information processing: the science system consumes, transforms, produces, and exchanges "information". Scientists talk to one another, read each other's papers,[7] and most importantly, they publish scientific papers. Scientometrics studies and analyzes, through the reference lists, the exchanges of information encoded in published papers. Scientists collect and analyze prior knowledge encoded in verbal forms, add value to it – producing new knowledge, which they nearly always encode in papers made accessible to other scientists, and so contribute to further scientific and technical advancement. For the production of new knowledge to have an impact "on scientific advancement", it has to be used by other scientists: no use, no impact. The same holds true for inventions: no commercialization, no innovation, no impact. Citations "certify" the use of prior knowledge.[8] We can conclude then that scientometrics measures the impact of research on future scientific and technical advancement, where publication is the research output, and citation is the natural indicator of its impact.

The underlying rationale is that, in order for a research result (as any goods and services) to have an impact it has to be "used". The question is whether and to what extent citations are certification of real use and representative of all use. This point is critical. It is about whether one believes that the norm is that scientists cite papers to recognize their influence, being aware that exceptions (uncitedness, undercitation, and overcitation) occur (Mertonian or normative theory of citing), or that the opposite is true, i.e. that citing to give credit is the exception, while persuasion is the major motivation for citing (social constructivism). Several overviews of the many studies in this area exist (Tahamtan,

---

[6] A literature review on how social impact can be assessed is provided by Bornmann (2013).
[7] By the terms paper, article, publication used in this work, I mean any document type (including patents) in which the new knowledge is encoded.
[8] Not in each and every case. See next for a discussion on the limits of evaluative citation analysis.

Safipour Afshar, & Ahamdzadeh, 2016; Tahamtan & Bornmann, 2018; Bornmann & Daniel, 2008). For a historical reenactment of the debate, I refer the reader to the beautiful homage paper presented by Harriet Zuckerman at the September, 2017 conference held in Philadelphia to celebrate Eugene Garfield's life. Suffice here to recall that Michael Mulkay (1976) and David Bloor (1976) were the first to develop the fundamentals of constructivism, followed by Gilbert (1977) and Latour (1987). More recently, the information scientist Terrence Brooks and biologists Michael and Barbara MacRoberts have contributed to keeping the debate alive, by objecting to the basic assumption of the normative theory of citing – i.e. that scientists cite their influences (Brooks, 1985, 1986; MacRoberts & MacRoberts, 1984, 1987, 1988, 1989a, 1989b, 1996, 2018).

I personally tend to agree with the Mertonian, normative conception of what citations signify (Kaplan, 1965; Merton, 1973) and how they function. This does not mean brushing aside all the criticisms of the above mentioned *social constructivists*. Still, a social or economic model is a proxy representation of entities and relationships between them, which captures the norm and inevitably leaves out the exceptions. In the case in point, stating that citations certify use, does not imply that there are no exceptions, rather that it is the norm. At scale, limits and exceptions then should not be an excuse for, and do not legitimate vague definitions of impact or a multitude of indicators for its measurement. Obviously, from an orthodox constructivist standpoint, evaluative scientometrics is nonsense and this manuscript would have no grounds. This paper addresses those who believe that the above limits are relatively small and therefore fail, at scale, to bias the information contained in the huge corpus of citations in the literature.

Finally, it must be noted that evaluative citation-based analysis is unable to capture impact outside the scientific system, such as that on practitioners (e.g. a physician applying a new pharmacological protocol after reading relevant literature), in education (scientists serving in the transmission of new knowledge to students) or on technology (e.g. scientists acting as consultants to industry and government). My conclusion, which is not new in the scientometric literature, is that evaluative scientometrics is aimed at measuring, or more precisely, predicting the impact of research on future scientific and technical advancement. Therefore, its most descriptive names are probably "scientific impact" or "scholarly impact", and not "social impact".

## 3. The measurement of research impact

### 3.1 Impact indicators

In evaluative scientometrics, all the usual limits, caveats, assumptions and qualifications apply, in particular: i) publications as not representative of all knowledge produced; ii) bibliometric repertories do not cover all publications; and iii) citations are not always certification of real use and representative of all use. Avoiding any possible confusion: in scientometrics, the basic unit of analysis of scholarly impact is the publication. It is the publication, i.e. the knowledge embedded therein, that can have a scholarly impact. The association of impact with other things, such as scientific journals, researchers, organizations, etc. could be more or less useful, but in any case is indirectly measured through the impact of the relevant publications.

How then to measure the scholarly impact of publications? We frequently hear that the proper basis of research evaluation would be for experts to review the work of their colleagues. Indeed peer-review, rather than metrics, is often used to evaluate the "quality" of publications. The United Kingdom's RAE and REF research assessment exercises are typical examples of peer-review evaluation exercises. But in fact reviewers cannot assess quality=impact, at the most they may be able to predict it, by examining the characteristics of the publication under evaluation. To better convey the concept, *mutatis mutandis*, the commercial success of a product is ultimately measured by sales, never by prior judgment of its market appeal. Now in scientometrics, we have seen that citation is the natural indicator of impact, as it certifies the use of the cited publication towards the scientific advancement encoded in the citing publication. Although a reviewer might judge certain characteristics of a publication, these characteristics cannot certify use. Also, no other indicator certifies use. The journal impact, whether measured by the journal's impact factor (IF), CiteScore, IPP, SNIP or other, reflects the distribution of citations of all hosted publications, not the individual ones. It simply designates the prestige of the journal. Returning to our business example, the revenue of a commercial venture is measured by the total amount of cash generated by the sale of products, not by the prestige of the location. Although we might expect to sell more from a shop on 5th Avenue than we would from one in the Bronx, in the end it is only the sales that will tell the bottom line. Altmetric.com scores, i.e. Mendeley reader counts, CiteULike, etc., again do not certify the scholarly use of a publication. Reading or downloading a publication are the same as, respectively, window shopping and trying on clothes. They do not certify a "buy", and therefore use. Other alternative indicators (news, blogs, twitter, "likes") offer even less "certification". However, as we will soon see, this does not mean that journal impact and altmetrics would have no use in assessing the scholarly impact of a publication.

Ideally, when the life of a publication is over, i.e. when it is no longer cited,[9] the accrued citations reflect the publication's impact. But reality poses difficult challenges for the scientometrician. For most practical purposes, impact assessment must be carried out not too long after publication. Policy-makers and managers, hoping to make informed decisions, cannot wait the many years (or decades) for citation life-cycles to end, and so to conduct research assessments. Still, the scientometrician should not complain, as the managers involved in the preceding stage of the value chain have a much more challenging task: selecting projects to execute, and establishing whether they are worth conducting, based on a more or less informed guess about yet-to-arrive results, outcomes, returns, technical and commercial risks.[10] Scientometricians find themselves in a much more comfortable position, dealing with known, finite products and their early entry in the "science market". The challenge is how to predict future impact, which is neither more nor less difficult than predicting future sales of some other product, newly launched on the market.

---

[9] "Obliteration by incorporation" or OBI is considered an exception, rather than the norm. Obliteration of the source of research results occurs when they become so well-known that scientists simply do not cite them. In any case, OBI reduces the rate of citations to already highly cited works, with negligible negative effects, as they already are highly cited. Merton first introduced the concept in his *Social Theory and Social Structure* in 1949 (although the revised edition of 1968 is usually cited, 27–28; 35–37, in the enlarged edition).

[10] To the purpose, a wide variety of project selection models have been assembled over the years, including linear programming, scoring models, checklists, up to the sophisticated R&D options selection models (Perlitz, Peske, & Schrank, 1999), which reveal more or less appropriate depending on the different types of research.

## 3.2 Predicting future impact

The questions for evaluative scientometrics are: i) which indicator (or combination) best predicts future impact; and ii) what is its predictive power. The answers are not univocal, as they depend on the time elapsed from date of publication to measurement of impact. What should not be in discussion is that late citation counts (as proxy of long-term impact) serve as the benchmark for determining the best indicator (and its predictive power), in function of the citation time window. Also clear is that the scientometrician must transmit, to the decision maker, the knowledge of the embedded tradeoff between level of accuracy and timeliness in measurement.

Initially, scholars faced the problem of how long the citation time window should be for the early citations to be considered as an accurate proxy of later citations. Adams (2005) states that citations accrued one or two years after publication "might be useful as a forward indicator of the long-term quality of research publications". Mingers (2008), Wang (2013), Baumgartner and Leydesdorff (2014) observe that the answers differ across disciplines. Rousseau (1988), and Glänzel, Schlemmer, and Thijs (2003) note that for mathematics the citation time horizon must be greater than for other fields. Abramo, Cicero, and D'Angelo (2011) found that in biology, biomedical research, chemistry, clinical medicine and physics, the peak in citations occurs in the second year after publication; in earth and space science and in engineering, citations follow a more regular and slower-growing trend. Mathematics behaves still differently, with publications collecting citations very slowly. Sugimoto and Larivière (2018) provide timelines for all the major scientific fields.

Scientometricians have investigated the possibility of increasing the accuracy of impact prediction by combining early citation counts with other independent variables. Virgo (1977) proved that a linear function of citation frequency and IF, determined through a multiple regression analysis, is a stronger indicator of the importance of an article, than citations alone. Through a mechanistic model collapsing the citation histories of publications from different journals and disciplines into a single curve, Wang, Song and Barabási (2013) concluded that "all papers tend to follow the same universal temporal pattern". Wang, Mei, and Hicks (2014) objected that "their own analyses find discouraging results… and correspondingly enormous prediction errors. The prediction power is even worse than simply using short-term citations to approximate long-term citations". Several years earlier, Abramo, D'Angelo, and Di Costa (2010) had provided proof that for citation windows of two years or less, the journal's IF is a better predictor of impact than citations, for articles in mathematics (and with weaker evidence in biology and earth sciences).

Levitt and Thelwall (2011) verified the predictive power of a combination of IF and citations, and recommended the hybrid indicator for time windows of zero or one year only, with the exception of mathematics, where its use for up to a two-year window has been suggested. Bornmann, Leydesdorff, and Wang (2014) also found that IF can be a significant co-variate in predicting the citation impact of individual publications. Stern (2014) confirmed that in the social sciences, IF improves correlation between the predicted and actual ranks by citations, when applied in the "zero" year of publication and up to one year afterwards. Stegehuis, Litvak, and Waltman (2015) proposed a model to predict a probability distribution for the future number of citations of a publication using the IF of the hosting journal and the number of citations received by the publication within one year of appearance. The latest Italian research assessment exercise (VQR 2011-2014) adopted a linearly weighted combination of citations and journal metric percentiles (Anfossi et al., 2016), but Abramo and D'Angelo (2016a) highlighted that citations alone showed a stronger predictive power than the combination adopted in the VQR. Abramo, D'Angelo, and Felici (2018) showed that the role of the IF in the prediction becomes negligible after only two years from publication. The authors suggest that in the future, scientometricians investigate the possibilities of using also altmetrics as covariates,[11] to improve predictive power when the citation time window is too short and early citations are too few.

Quantitative studies on the recourse to the expensive and time consuming option of peer-review have been limited to measuring the correlation between peer-review and bibliometric evaluations: i) of publications (Bertocchi et al., 2015; Reale, Barbara, & Costantini, 2006; ); ii) of individual researchers or groups (Van Raan, 2006; Aksnes, & Taxt, 2004; Oppenheim, & Norris, 2003; Meho, & Sonnenwald, 2000; Rinia, Van Leeuwen, Van Vuren, & Van Raan, 1998; Oppenheim, 1997); and iii) of research institutions (Franceschet, & Costantini, 2011; Abramo, D'Angelo, & Di Costa, 2011; Thomas, & Watkins, 1998). The limits of many of these studies are that the correlation has been demonstrated only for a limited number of scientific fields and/or that the quantitative indicators used for the comparison have not been appropriate. In any case, none of these analyses resolve the dilemma of which methodology, peer-review vs early citations, would be the more precise and reliable, in function of the citation time window, to predict the long-term impact of publications.

## 4. Conclusions

As with the concept of research performance (Abramo & D'Angelo, 2014) and its measurement (Abramo & D'Angelo, 2016b; 2016c), I feel the same urgency to continue a lively discussion that has deep roots in the field, aiming at definitive convergence on the meaning and measurement of the concept of impact, a concept that I regard as key to scientometric science. While it is true that doubts and curiosity are drivers of scientific advancement, ambiguities should not be allowed to weaken the foundations of a discipline.

This conceptual work attempts to spell out some principles leading to a clear definition of "research impact" and a proposal for the appropriate scientometric indicator to measure it. From the citation-based evaluative scientometric perspective, research impact means "the contribution of research output to further scientific and technical advancement", which we call "scientific or scholarly impact", since citation-based evaluative scientometrics does not deal with social impact. Making this clear should assist decision makers in understanding what they can truly expect from evaluative scientometrics. For example, we frequently read that national research assessment exercises serve in demonstrating the delivery of public benefits from research investment, when in fact scientometrics is incapable of such demonstration. What we could perhaps do is demonstrate cases of positive correlations between scholarly impact and public benefits, as it has been shown for citation rates of patents and their market value, whereby an extra

---

[11] Currently, records of altmetrics data are too recent to serve such purposes.

citation per patent boosts market value by 3%, according to Hall, Jaffe, and Trajtenberg (2005).

Citation count is the natural indicator of scholarly impact, in that a citation certifies the use of the knowledge encoded in the cited publication. In carrying out impact measurement, the many caveats of evaluative citation analysis still apply. Applied scientometrics has to deal with the embedded tradeoff between level of accuracy and timeliness in impact measurement, for evaluation purposes. However, when the citation time window is too short, alternative indicators, alone or in combination with citations, can increase the predictive power and accuracy of impact measurement. The identification of alternative indicators or combinations, which could vary in type and weight across disciplines, as a function of the citation time window, and early-citations accrued, should be conducted by comparing their prediction of impact to the benchmark of late citation counts.

I do not expect that my perspectives on the conceptualization of impact will be agreeable to all, nor my views on measurement. I hope that criticisms, qualifications, and suggestions will be spelled out, so that the scientometric epistemic community can move toward a shared definition of research impact, and identification of the most appropriate indicator for its measurement. While a shared definition and indicator for measurement are helpful in making better elaborations, conducive of more informed decisions, users are warned to keep in mind all the limits, caveats and assumptions of evaluative scientometrics.